\documentstyle[twocolumn,prl,aps]{revtex}
\begin{document}
\title{Critical fields and devil's staircase in superconducting ladders}
  
\author{ Richard T Giles and Feodor V Kusmartsev }
\address{Department of Physics, Loughborough University, Loughborough, LE11
3TU, U.K.}
\date{\today}
\maketitle

\begin{abstract}
  We have determined the ground state for both a ladder array of
  Josephson junctions and a ladder of thin superconducting wires.  We
  find that the repulsive interaction between vortices falls off
  exponentially with separation. The fact that the interaction is
  short-range leads to novel phenomena. The ground state vortex
  density exhibits a complete devil's staircase as the applied
  magnetic field is increased, each step producing a pair of
  metal-insulator transitions. The critical fields in the staircase
  are all calculated analytically and depend only on the connectivity
  of the ladder and the area of the elementary plaquette.  In
  particular the normal square ladder contains no vortices at all
  until the flux per plaquette reaches $~\frac{\Phi_0}{2\sqrt{3}}$.
\end{abstract}

\pacs{05.50.+q, 74.50.+r, 74.60.Ec, 85.25.-j(Kx,Cp)}

It is now relatively easy to fabricate complex structures containing
many Josephson junctions. Two-dimensional arrays have long been of
interest.  In a transverse magnetic field, both square\cite{Wees} and
triangular\cite{Ther,Zant,Ther2} 2D arrays show very rich structure in
resistance, impedance and inverse sheet inductance measurements when
the flux per plaquette is close to a simple rational fraction (at
which commensurate states form).  Such arrays appear to be well
described by a frustrated XY
model\cite{Teit,Shih,Choi,Minn,Gran2,Shen,Geig,Ario,Gran} although
special attention has only been given\cite{Teit,Gran2,Li} to the
simplest rational flux values: $f = \frac{1}{2}, \frac{1}{3}$ and
$\frac{2}{5}$.  Vortex dynamics in 2D arrays display many anomalous
features \cite{Zant2,Jons}. Interesting
self-inductance\cite{Phil,Stra} and charging\cite{Zant3} effects are
also observed.

Analogous phenomena are observed in superconducting weakly coupled
wire networks made in thin amorphous niobium silicon films
\cite{Zant4,Jean}. Furthermore Monte-Carlo calculations\cite{Fran} on
a Coulomb gas model of a 2D network have indicated the existence of
two melting transitions in the vortex sub-system.

Recently there has been much interest in ladder arrays of Josephson
junctions.  Experimental work\cite{Oude} on wide ladders has found a
rich structure in the dependence of the longitudinal resistance on
transverse magnetic field and this has been interpreted in terms of
metal-insulator transitions. In numerical calculations\cite{Himb} it
has been noticed that the normal square Josephson ladder exhibits a
critical field below which the ground state contains no vortices while
Monte-Carlo simulations have shown, for the case $f=\frac{1}{2}$, that
the ladder belongs\cite{Gran3} to the universality class of the Ising
model.

A quasi-classical analytical method for investigating Josephson
networks has been successfully tested\cite{Faruque} against full
Monte-Carlo simulations.  In this paper we use the same method to
completely determine the ground state, as a function of applied
magnetic field, of both the Josephson ladder shown in Fig.~1a and of
the analogous superconducting wire network shown in Fig.~1b. We find
that the vortex-vortex interaction falls off exponentially with
separation, in contrast to the long-range logarithmic interaction
between vortices in 2D and 3D arrays, and that at low temperatures the
ladder is described by an Ising model. The short-range nature of the
interaction gives rise to a geometry dependent critical field $H_g$
which depends solely on the geometrical structure of the ladder and
plays the same role as $H_{c1}$ in bulk superconductors. Above $H_g$
the vortex density exhibits a complete devil's staircase as the
applied field is increased.  To our knowledge, this is the first
example of a real physical system exhibiting a non-trivial complete
devil's staircase.

First we study the Josephson ladder and then show that the analogous
ladder of superconducting wires behaves in the same way. A network of
Josephson junctions can be described by the XY-model\cite{Orlando}:
\begin{equation}
  H = \frac{I_c\Phi_0}{2\pi}\sum_{i}(1-\cos\phi_{i})
\end{equation}
where the sum is over all junctions, $I_c$ is the critical current and
$\Phi_0=\frac{h}{2e}$ is the flux quantum. The gauge-invariant phase
difference $\phi_i$ across the $i$th junction is defined by
\begin{equation}
  \phi_{i}=\Delta\theta_i - 
           \frac{2\pi}{\Phi_0}\int_i \mathbf{A}.\mathrm{d}\mathbf{l}
\end{equation}
where $\Delta\theta_i$ is the phase difference and the line integral of the
vector potential $\mathbf{A}$ is taken across the junction. The
current through the $i$th junction is given by
\begin{equation}
  \label{eq:I}
  I_i=I_c\sin\phi_i
\end{equation}

Consider a ladder (Fig.~1a), which we call the Josephson ladder,
consisting of a linear series of identical plaquettes. Let there be
$2u$ unshared junctions in each plaquette and $c$ junctions in each
region which is in common between two plaquettes.  Let the applied
magnetic flux per plaquette be~$f\Phi_0$.  The sum of the gauge
invariant phase differences around any plaquette is then quantised
according to:
\begin{equation}
  \label{eq:gauge}
  \sum_{\mathrm{plaquette}\ \mathit{j}}\frac{\phi_{i}}{2\pi}
 =n_j-f-\frac{1}{2(u+c)I_{\Phi_0}}\sum_{\mathrm{plaquette}\ \mathit{j}} I_i
\end{equation}
where $n_j$ is the vortex number and $I_{\Phi_0}$ is the current
which, if circulating around a plaquette, would give an induced flux
of one flux quantum~$\Phi_0$.  The last term gives a good
approximation to the induced flux if, for example, the junctions are
uniformly spaced around a rectangle.

If $u\neq 0$ and if the number of junctions in each
plaquette is large then the phase $\phi$ dropped across each
junction will be small enough that we have
$\sin\phi\approx\phi$ and so Eqs.~(\ref{eq:I},\ref{eq:gauge}) give us
\begin{equation}
  \sum_{\mathrm{plaquette}\ \mathit{j}} I_iR =n_j-f
  \label{eq:Kirchoff}
\end{equation}
where the `resistance'~$R$ is
\begin{equation}
  \label{eq:R}
  R = \frac{1}{2\pi I_c}+\frac{1}{2(u+c)I_{\Phi_0}}
\end{equation}
In this approximation the energy of junction $i$ becomes
$I_i^2\Phi_0/(4\pi I_c)$ and so the total energy $U$ is
\begin{equation}
  \label{eq:Udef}
  U= \kappa \sum_{\mathrm{all\ junctions}} I_i^2
\end{equation}
where $\kappa=\Phi_0/(4\pi I_c)$.
Since the currents are linear in $f$ the energy $U$
must be quadratic in $f$.

Now we consider a similar ladder of superconducting wires as shown in
Fig.~1b. $ua$ and $ca$ are the lengths of the wires, $u$ and $c$ being
integers. When the
true superconducting state is established the superconducting order
parameter takes the form
$\psi=\psi_0 \exp\left\{i\left(\theta-\frac{2\pi}{\Phi_0}\int
    \mathbf{A}.\mathrm{d}\mathbf{l}\right)\right\}$.
If the width of each wire is small compared with the London
    penetration depth then vortices cannot
    penetrate the wires; if in addition the width is small compared with
    the length then the current density~$J$ will be constant:
\begin{equation}
  \label{eq:curr}
J = \frac{e|\psi_0|^2\hbar}{m}\left|\nabla\theta
  -\frac{2\pi}{\Phi_0}\mathbf{A}\right|\equiv \frac{I}{S}
  \end{equation}
  where $S$ is the wire cross-sectional area and $I$ is the current
  carried by the wire.  Integrating around a plaquette we again obtain
  Eq.~(\ref{eq:Kirchoff}) but now we have
\begin{equation}
  \label{eq:Rwire}
  R= \frac{ma}{2\pi e\hbar|\psi_0|^2 S} + \frac{1}{2(u+c)I_{\Phi_0}}
\end{equation}
The total energy of
the ladder is likewise given by Eq.~(\ref{eq:Udef}) except that now
$\kappa=m/(4e^2|\psi_0|^2 S)$.

Thus Eqs.~(\ref{eq:Kirchoff}) and (\ref{eq:Udef}) determine the
solution of the currents and total energy in both the Josephson ladder
and a ladder of long thin superconducting wires (and also, of course,
the solution of the currents and power dissipation in a similar
network of resistors and batteries).  This method has already been
tested\cite{Faruque} against Monte-Carlo calculations for a number of
small Josephson networks.

Solving Eqs.~(\ref{eq:Kirchoff}) and (\ref{eq:Udef}) for a ladder of
infinite length we find that the energy $U$ is
\begin{equation}
  \label{eq:U}
  \frac{U}{2U_0} = \sum_j \varepsilon_j^2
      + 2\sum_j \sum_{k>j} \alpha^{k-j}\varepsilon_j\varepsilon_k
\end{equation}
where $\varepsilon_j=n_j-f$,
$U_0=\kappa/(4R^2\sqrt{u^2+2uc})$ and
\begin{equation}
  \label{eq:alpha}
  \alpha=1+\frac{u}{c}-\sqrt{\frac{2u}{c}+\frac{u^2}{c^2}}
\end{equation}
$\alpha$ depends solely on $u/c$ as shown in Fig.~2; all
other factors contribute only to the energy scale~$U_0$. Note that
vortices repel each other ($\alpha<1$) and since the energy
decreases exponentially with separation the vortex-vortex interaction
is a short-range interaction of range $-1/\ln\alpha$ plaquettes.

We now focus on finding the lowest energy state for any given value of
$f$. For $f$ in the range $0<f<1$ we need only consider two values of
$n_j$, namely $n_j=0$ and $n_j=1$.  Hubbard\cite{Hubbard} has shown
that for particles (in our case vortices, i.e. plaquettes with
$n_j=1$) interacting with a convex repulsive interaction the lowest
energy state of $p$ particles in a one-dimensional chain on $q$ sites
is a generalised Wigner lattice consisting of a sequence of two
spacings, the spacings being the two integers which bracket $q/p$ (for
example, when $q/p=7/3$ the two spacings will be 2 and 3).  Fig.~3
shows the energies of these generalised Wigner lattices for various
vortex densities $\rho=p/q$ as calculated by direct solution of
Eq.~(\ref{eq:Kirchoff}).  Immediately we see that there is a critical
flux~$f_c$ below which the empty ladder (i.e. $\rho=0$) is the lowest
energy state.  A more accurate numerical solution was performed for
the same Josephson ladder, $\phi_i$ in Eq.~(\ref{eq:gauge}) being
replaced by $\sin^{-1}(I_i/I_c)$ instead of $I_i/I_c$; this more
accurate calculation gave the transitions from one generalised Wigner
lattice to another as occurring at much the same values of $f$,
differing by no more than $~0.01$ from the values seen in Fig.~3. A
ladder of larger $u$ and $c$ would have shown better agreement.

We can calculate the critical flux~$f_c$ very simply.  The energy
increase $\delta U$ when one vortex is added to an empty ladder is
\begin{equation}
  \label{eq:single}
    \frac{\delta U}{2 U_0} =  1-2\frac{1+\alpha}{1-\alpha}f 
\end{equation}
Thus $\delta U>0$ for $f<f_c$ where
\begin{equation}
  \label{eq:fcrit}
  f_c=\frac{1-\alpha}{2(1+\alpha)}
\end{equation}
i.e. for $f<f_c$ the lowest energy state is the empty ladder.
Furthermore we see that, as shown in Fig~2, the critical flux $f_c$
depends solely on the ladder ratio $u/c$; it is independent of  all other
parameters. Eq.~(\ref{eq:fcrit}) is consistent with the result
obtained previously\cite{Himb} for the square ladder with nearest
neighbour interactions only.

It may look as though Fig 3 not only tells us the ground state but
also the low energy excited states.  However this is not the case.
Excited states are not necessarily generalised Wigner lattices; in
fact since the interaction is short range the lowest energy excitation
will normally be a domain wall.

Since, for low energy states, there are only
two possible values of $\varepsilon$ (i.e. $\varepsilon=-f$ and
$\varepsilon=1-f$) Eq.~(\ref{eq:U}) is reminiscent
of the Ising model Hamiltonian.
Let
\begin{equation}
  \label{eq:sdef}
  s_j=\left\{ \begin{array}{ll}
                -1 & \mbox{if $n_j=0$ } \\
                +1 & \mbox{if $n_j=1$ } 
              \end{array} \right.
\end{equation}
Re-writing Eq.~(\ref{eq:U}) in terms of these Ising variables we
obtain for a ladder of $N$ plaquettes (valid in the limit
$N\rightarrow\infty$)
\begin{eqnarray}
  \frac{U}{U_0}& = & \frac{N}{2}
  +\frac{\left(\frac{1}{2}-f\right)^2 N}{f_c} \nonumber \\
   & & \mbox{}+ \frac{\left(\frac{1}{2}-f\right)}{f_c}
        \sum_j s_j + \sum_j\sum_{k>j}\alpha^{k-j} s_j s_k
  \label{eq:ising}
\end{eqnarray}

This is an anti-ferromagnetic Ising model Hamiltonian in which
$\frac{1}{2}-f$ plays the role of an applied field, the
interaction between spins falling off exponentially with separation.
The fact that at low temperatures the ladder is described by
an Ising model Hamiltonian explains the Monte-Carlo result\cite{Gran}
(determined only for the
special case $f=1/2$) that the Josephson ladder belongs to
the universality class of the Ising Model.

We now explore the implications of Eq.~(\ref{eq:ising}). Published
literature\cite{Baker,Stephenson,Shrock} gives no complete solution of
the thermodynamics of this kind of Ising model at both non-zero field
and non-zero temperature.  It has been shown\cite{Bak} however that
the ground state of a one-dimensional Ising model with an
anti-ferromagnetic interaction exhibits a complete devil's
staircase\cite{Mandelbrot} if the interaction is convex (i.e.
decreases with separation with positive 2nd derivative); \textit{the
  average spin~$\left<s\right>$ per site is always a rational
  fraction, all rational fractions between 0 and -1 being visited as
  the field (in our case $\frac{1}{2}-f$) is increased from zero to
  the critical field (in our case $\frac{1}{2}-f_c$) above which the
  spins are all aligned opposite to the field (in our case the ladder
  contains no vortices)}.

Krantz \textit{et al}\cite{Krantz} have calculated the precise nature
of the expected devil's staircase for the case of an
anti-ferromagnetic Ising model with an exponential interaction between
spins.  Using Eq.~(\ref{eq:ising}) their results may now be applied to
the superconducting ladder. Fig.~3 shows the predicted devil's
staircase for the `square' ladder (i.e. $u=c=1$); note that the
locations of the transition points are seen to agree with the crossing
points of the energy curves. A ladder with $u<c$ would give a
staircase in which steps corresponding to larger values of the
denominator $q$ would be more easily resolved.

In conclusion, we have shown that in the Josephson ladder and in the
superconducting wire ladder \textit{the interaction between vortices
  falls exponentially with separation}; this is in contrast to the 2D
array, where the logarithmic interaction between vortices leads to the
Kosterlitz-Thouless phase transition\cite{KT}.  At low temperatures
\textit{the ladder is described by a 1D~Ising model}
(Eq.~(\ref{eq:ising})) and so no phase transition is
expected\cite{Ziman} except at zero temperature.  This also leads to
the \textit{ground state vortex density exhibiting a complete devil's
  staircase as the applied magnetic field is increased}; at each step
in the staircase one extra vortex penetrates the ladder. To our
knowledge this is the first example of a physical system exhibiting a
non-trivial devil's staircase. The fact that the devil's staircase
occurs in both the Josephson ladder and the wire ladder suggests that
it is a very general property of superconducting ladder structures.

In 2D networks and in wide multi-leg ladders\cite{Oude} commensurate
states have been observed centred on rational flux values; note that
our results suggest that this is \emph{not} the case for the 1D
ladder, e.g.  a vortex density of $\rho=1/5$ doesn't occur at $f=1/5$
but in a small region close to $f=0.3$. Likewise, the critical field
$f_c$ has no apparent analogue in 2D. These important differences
arise because in the simple 1D ladder the interaction is short-range
(exponential) whereas in 2D it is logarithmic.  It is interesting that
in the long range limit (i.e. small $u/c$) the properties of the 1D
ladder are reminiscent of the 2D ladder: i.e.  $f_c\rightarrow 0$ and
the ground state vortex density $\rho=p/q\rightarrow f$.

We call for experiments to detect the predicted devil's staircase
phenomenon which should be observable in low temperature mutual
inductance and resistivity measurements. Each step in the staircase
should give rise to a pair of metal-insulator transitions\cite{Oude}:
close to a step the disorder will cause high resistance while far away
from any step the ordered array of vortices will cause low resistance.

From a practical point of view perhaps the most important result is
that \textit{both kinds of ladder exhibit a geometry-dependent
  critical field~$H_g$}, analagous to $H_{c1}$, below which all
vortices are expelled. Applying Eqs.~(\ref{eq:alpha})
and~(\ref{eq:fcrit}) one can design a ladder of any desired
$H_g=f_c\Phi_0/A$ by appropriately choosing the ladder ratio $u/c$ and
the area~$A$ of the elementary plaquette.

We would like to thank John Samson and Daniil Khomskii for
illuminating discussions.

\begin{figure}
\caption{Various ladder structures: 
  (a) the general Josephson ladder, (b) the wire ladder.}
\end{figure}
\begin{figure}
\caption{The dependence of $f_c$ and $\alpha$ on the ladder ratio $u/c$.}
\end{figure}
\begin{figure}[htbp]
  
  \caption{The solid curves show the dependence of energy per
    plaquette~$U/(NU_0)$ on flux per plaquette~$f$ for the `square'
    ladder, i.e. $u=c=1$; each curve represents a generalised Wigner
    lattices of a particular vortex density~$\rho=p/q$. The energies are calculated for a 512 plaquette ladder by direct
    solution of Eqs.~(\ref{eq:Kirchoff}) for the case
    $I_{\Phi_0}=\infty$. The broken line shows the predicted devil's
    staircase in the ground state vortex density; each visible step
    contains an infinite number of unresolved steps. Note the critical
    flux at $f=f_c=1/(2\sqrt{3})\approx 0.289$.}
  \label{fig:Parabolae}
\end{figure}

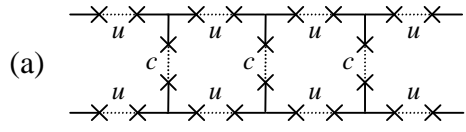
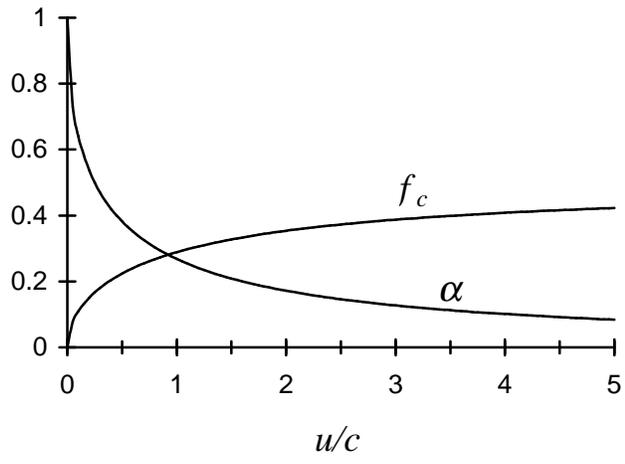
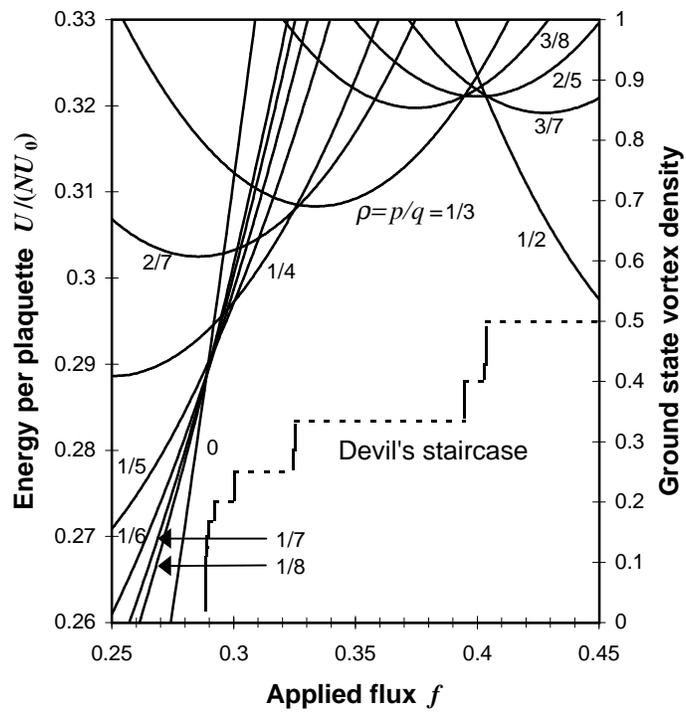
\end{document}